\documentclass{WileyMSP-template}

\usepackage{gensymb}
\usepackage{textgreek}
\usepackage{textcomp}
\usepackage{graphicx}
\usepackage[export]{adjustbox}
\usepackage[all]{nowidow}
\usepackage[labelsep=period]{caption}
\usepackage{mathtools}
\usepackage[T1]{fontenc}
\usepackage{ragged2e}

\begin{document}

\pagestyle{fancy}

\title{3D Nanoscale Mapping of Short-Range Order in GeSn Alloys}

\maketitle


\author{Shang Liu}
\author{Alejandra Cuervo Covian}
\author{Xiaoxin Wang}
\author{Cory T. Cline}
\author{Austin Akey}
\author{Weiling Dong}
\author{Shui-Qing Yu}
\author{Jifeng Liu*}


\dedication{}

\begin{affiliations}
Shang Liu, Alejandra Cuervo Covian, Xiaoxin Wang, Cory T. Cline, Weiling Dong, Jifeng Liu\\
Thayer School of Engineering, Dartmouth College, 14 Engineering Drive, Hanover, NH 03755, USA\\
Email Address: Jifeng.Liu@dartmouth.edu\\
\medskip
Austin Akey\\
Center for Nanoscale Systems, Harvard University, Cambridge, MA 02138, USA\\
\medskip
Shui-Qing Yu\\
Department of Electrical Engineering, University of Arkansas, Fayetteville, AR 72701, USA

\end{affiliations}



\setlength\parindent{24pt}
\justifying

\begin{abstract}

GeSn on Si has attracted much research interest due to its tunable direct bandgap for mid-infrared applications. Recently, short-range order (SRO) in GeSn alloys has been theoretically predicted, which profoundly impacts the band structure. However, characterizing SRO in GeSn is challenging. Guided by physics-informed Poisson statistical analyses of Kth-nearest neighbors (KNN) in atom probe tomography, a new approach is demonstrated here for 3D nanoscale SRO mapping and semi-quantitative strain mapping in GeSn. For GeSn with $\sim$14 at.\% Sn, the SRO parameters of Sn-Sn 1NN in 10x10x10 nm$^{3}$ nanocubes can deviate from that of the random alloys by $\pm$15 \%. The relatively large fluctuation of the SRO parameters contributes to band-edge softening observed optically. Sn-Sn 1NN also tends to be more favored towards the surface, less favored under strain relaxation or tensile strain, while almost independent of local Sn composition.  An algorithm based on least square fit of atomic positions further verifies this Poisson-KNN statistical method. Compared to existing macroscopic spectroscopy or electron microscopy techniques, this new APT statistical analysis uniquely offers 3D SRO mapping at nanoscale resolution in a relatively large volume with millions of atoms. It can also be extended to investigate SRO in other alloy systems.

\end{abstract}


\section{Introduction}

Structure-property relationship has been the central theme of materials science. Of different hierarchies of materials structures, short range order (SRO) of atoms is an important factor bridging amorphous and crystalline materials, as well as random and ordered alloys. Recently, SRO in high and medium entropy metallic alloys have attracted great attention due to its impact on mechanical properties.\cite{zhang2020short,chen2021direct} Interestingly, SRO has also been predicted recently for GeSn covalent semiconductor alloys,\cite{cao2020short} a promising candidate for mid-infrared optoelectronics on Si \cite{al2016optically,elbaz2020reduced} due to their tunable direct bandgap \cite{homewood2015rise,soref1991predicted,alberi2008band} and compatibility with Si-based integrated circuits.\cite{moutanabbir2021monolithic,wang2019gesn} Theory has predicted \cite{cao2020short} that GeSn alloys with >10 at.\% Sn composition are not homogeneous random solid solutions but rather have SRO in certain preferences of atomic species for the Kth-nearest neighbors (KNN, K=1,2,3..), which profoundly impacts their band structures. Particularly, GeSn alloys of the same composition and lattice constant but different SRO may result in different bandgaps, opening an intriguing path towards perfectly lattice-matched GeSn SRO heterojunctions.

However, experimentally characterizing SRO is challenging. While extended X-ray absorption fine structure (EXAFS) has been applied to investigate the averaged local environment in Ge-rich GeSn alloys on macroscopic area,  very little information can be derived for Sn-Sn vs. Sn-Ge 1NN coordination numbers because the peak of the former is buried inside the dominant peak of the latter.\cite{gencarelli2015extended,robouch2020atomic} Indeed, the error bar of Sn-Sn 1NN coordination number is too large to tell if there is any significant deviation from random alloys \cite{robouch2020atomic} (also see Supporting Information). The corresponding 2NN coordination number error bars are also large, on the order of $\pm$1.\cite{gencarelli2015extended,robouch2020atomic} It has been hypothesized that different local configurations of Sn atoms may have led to the weak 2NN and 3NN signals,\cite{gencarelli2015extended} yet EXAFS itself is unable to probe these local configurations at nanoscale. This is further complicated by the lack of \textalpha-Sn foil as a reference for Sn-edge analyses, and the background signals from relatively thick Ge buffer layers for Ge-edge analyses. Advanced transmission electron microscopy (TEM) has been applied to study the SRO in medium entropy metallic alloys, yet each analysis is limited to a nm scale region \cite{zhang2020short,chen2021direct} such that it is hard to analyze a large number of nanoscale regions in TEM for more statistically conclusive results. 

In principle, atom probe tomography (APT) is a powerful experimental technique to study SRO by offering 3D atomic reconstruction of the alloy. Radial distribution function (RDF),\cite{philippe2010clustering,haley2009influence,mukherjee2017short,zhou2013quantitative} KNN,\cite{philippe2009clustering,stephenson2007new,tang2015indium,kumar2015interplay,gault2016brief} binomial distribution,\cite{moody2008quantitative,assali2018atomically} auto-correlation \cite{vurpillot2004application} and random labelling \cite{moody2007contingency,marceau2010evolution,ceguerra2010three} have been used to study the clustering in APT. In reality, however, high-fidelity retrieval of lattice information and accurate RDF has been limited to metals.\cite{gault2010spatial,geiser2007spatial} Deriving SRO of semiconductors alloys from APT has been a significant challenge due to larger measurement error in atomic positions compared to metals. The strong local magnification effect caused by non-uniform electric field during the field evaporation of semiconductors leads to higher or lower local atomic density.\cite{waugh1976investigations,birdseye1974analogue,philippe2010clustering} Consequently, the APT data of semiconductor alloys resembles a significantly perturbed lattice, and $\sim$70\% of the atomic sites are empty due to the limited detection efficiency. Therefore, one can no longer directly identify the lattice from the raw APT data. Currently, SRO analyses in alloys are limited to relatively crude methods such as artificially cutting a line midway between the theoretical distances of the k th and (k+1) th nearest neighbor shells to accommodate the perturbed atomic positions.\cite{ceguerra2012short,gault2012atom} 

To overcome this limit and retrieve SRO from APT data at high fidelity, here we demonstrate a new method that only requires the input of crystal structure to derive SRO in GeSn alloys from the nonideal APT data by correcting and refining KNN analysis with physics-informed statistical methods. This approach establishes a statistical correlation between the nominal KNN directly obtained from the nonideal APT data to the \emph{true} KNN shell for diamond cubic structure, thereby regrouping the former to reconstruct the latter at high fidelity. Lattice constants can also be derived using the peak positions of true KNN shell, allowing for semi-quantitative local strain mapping at nanoscale resolution in addition to SRO mapping. For GeSn alloys, we define SRO parameter $\alpha^{KNN}_{Sn-Sn}$ as the probability for a given Sn atom to have another Sn atom in its true KNN shell divided by the Sn at.\% in the same region. For example, $\alpha^{KNN}_{Sn-Sn}$>1 indicates more preference for a Sn atom to have other Sn atoms in its KNN shell than a random alloy. The preference of Sn-Sn 1NN also indicates a tendency of Sn-Sn clustering. Even though all species could be artificially ``clustered'' or ``de-clustered'' in terms of local atomic density due to strong local magnification effect in APT measurements of semiconductors, the SRO parameter defined this way represents the relative preference of Sn-Sn KNN normalized to the local Sn concentration in the \emph{same} region, thereby removing the artefact due to local atomic density fluctuations in APT and well representing the intrinsic SRO of the GeSn sample itself. 

Using this physics-informed statistical method, we are able to map the SRO distribution in 3D at $\sim$10 nm resolution in the GeSn thin films. Our study shows that $\alpha^{1NN}_{Sn-Sn}$ fluctuates between 0.85 to 1.15 in different regions. This  Sn-Sn 1NN SRO fluctuation in GeSn, together with the theoretical prediction of large band structure variation with SRO, contributes to the bandedge softening compared to pure Ge, as commonly observed in GeSn absorption spectra as well as  photoluminescence (PL) peak width$>>k_BT$ thermal broadening even at 10 K.\cite{olorunsola2021impact,zhou2020electrically,ye2015absorption,chen2014structural} The mean value of $\alpha^{1NN}_{Sn-Sn}$ decreases from 1.052$\pm$0.051 to 1.026$\pm$0.045 as the depth increase from 150 to 800 nm under the surface, with 18 10 nm $\times$10 nm$\times$10 nm nanocubes examined at each depth. These results indicate more preference towards Sn-Sn clustering near the surface, consistent with the tendency of Sn surface segregation. Furthermore, we found that Sn-Sn 1NN is generally less favored in strain-relaxed or tensile strain regions than compressive strained regions, while it is almost independent of the local Sn composition. We also found that, statistically, Sn-Sn 1NN-4NN are all slightly more preferred than those in random alloys, as opposed to slightly less favored Sn-Ge 1NN-4NN. These results qualitatively agree with previous EXAFS analyses in Ref.\cite{gencarelli2015extended} showing Sn-Sn 2NN is more preferred over Sn-Ge 2NN. To further verify this physics-informed statistical approach, we also developed an algorithm based on least square fit to rearrange atoms to nearby perfect lattice sites, which shows a very good agreement with the aforementioned SRO results from statistical methods. This new statistical APT analysis of SRO uniquely offers 3D SRO mapping and semi-quantitative local strain mapping at nanoscale resolution in a relatively large volume with millions of atoms (e.g., 30$\times$30$\times$800 nm$^3$). It nicely bridges the gap between macroscopically averaged EXAFS analyses and atomic scale STEM analyses in studying SRO, and it can be readily extended to other material systems such as high-entropy alloys.

\section{Methods and Results}

\subsection{APT data}
\begin{figure}[!htb]\centering
    \center{\includegraphics[width=\textwidth]
    {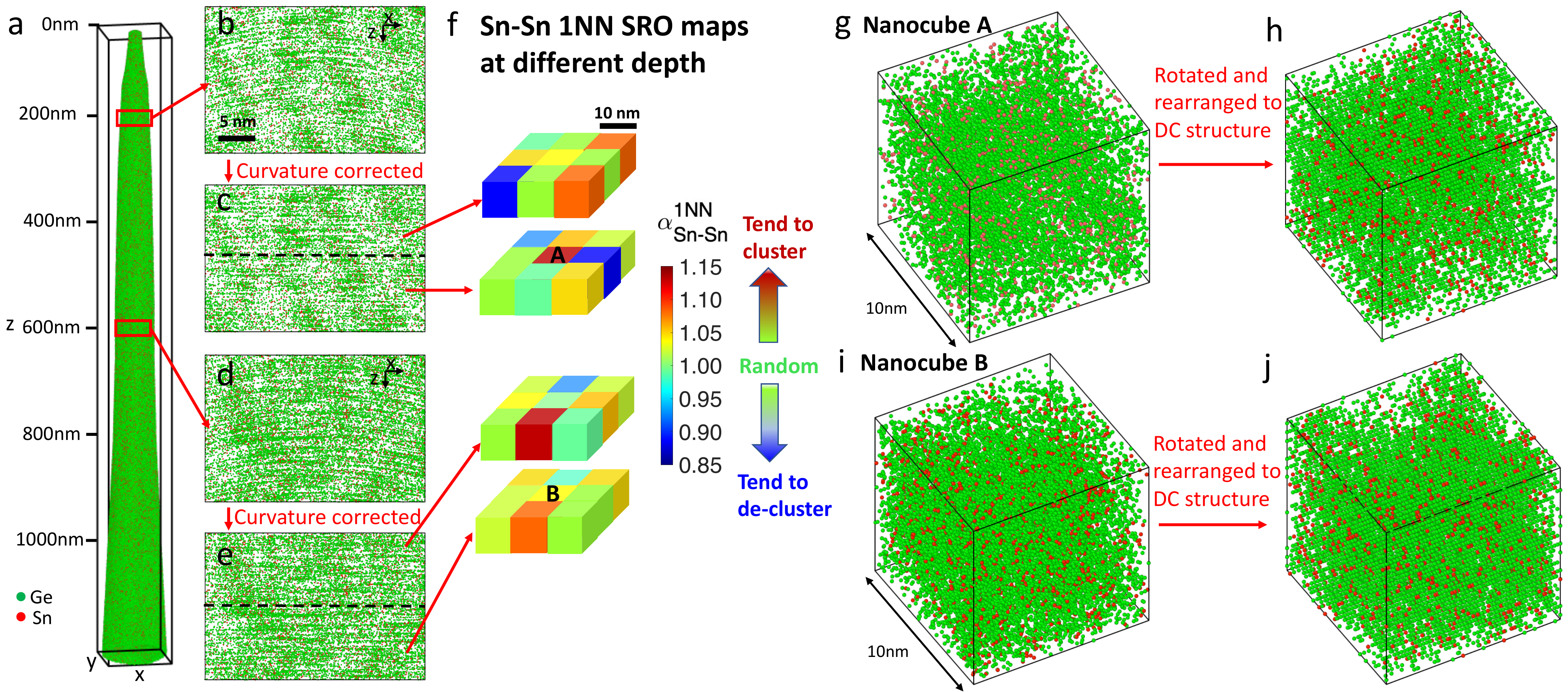}}
    \caption{(a) Reconstructed atomic map of a GeSn tip. (b), (d) Zoomed-in regions around z=200 nm and z=600 nm. (c), (e) Post-reconstruction curvature correction on (b) and (d). (f) $\alpha^{1NN}_{Sn-Sn}$ maps around z=200 nm and z=600 nm, where $\alpha^{1NN}_{Sn-Sn}$ is calculated for each 10 nm×10 nm×10 nm nanocube and color-labelled. (g), (i) Curvature corrected atomic maps of nanocubes A and B labeled in (f). (h), (j) Lattice rotation (by 45\degree around z-axis) and atomic rearrangement to diamond cubic (DC) structure with x, y and z axes aligned to <100> directions for nanocube regions A and B. Note that the visual differences before and after atomic rearrangement are mainly due to the lattice rotation, i.e. projecting the lattice in different perspectives.}
    \label{fig:Summary}
\end{figure}

\begin{figure}[!htb]\centering
    \center{\includegraphics[width=\textwidth]
    {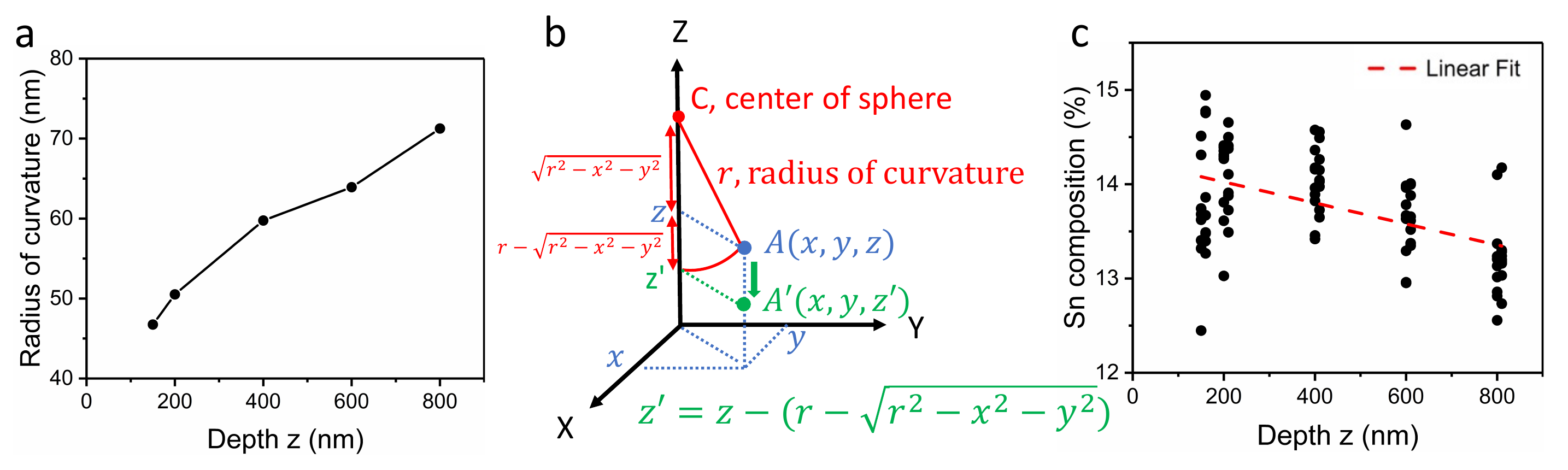}}
    \caption{(a) Radius of curvature of the reconstructed atomic map of the GeSn tip. (b) Schematics of post-reconstruction curvature correction. (c) Sn composition of 90 10 nm×10 nm×10 nm nanocubes after curvature correction.}
    \label{fig:CurvatureCorrection}
\end{figure}

GeSn films with $\sim$14 at.\% Sn are grown on relaxed Ge buffered Si substrate by chemical vapor deposition (CVD).\cite{dou2018investigation} A CAMECA LEAP 4000X HR 3D atom probe microscope is used for atom identification and spatial positioning. The APT data are reconstructed using IVAS software and the reconstructed atomic map of a GeSn tip is shown in \textbf{Figure \ref{fig:Summary}}(a). As shown in the zoomed-in images in Figure \ref{fig:Summary}(b) and (d), the lattice planes in reconstructed atomic map are not flat, so a post-reconstruction curvature correction \cite{vurpillot2011pragmatic,prosa2016approaches,vurpillot2013reconstructing,larson2011toward} is required. The relation between the radius of curvature of the lattice plane and the depth is plotted in \textbf{Figure \ref{fig:CurvatureCorrection}}(a): the further away from the top of the tip (i.e. the sample surface), the less accurate the reconstruction is. To flatten a lattice plane with a radius of curvature, $r$, the atom $A(x,y,z)$ on the plane should be moved to $A^{'}(x,y,z^{'})$ with
\begin{equation}
  z^{'} = z - (r - \sqrt{r^2-x^2-y^2}) ,
  \label{eqn:CurvatureCorrection}
\end{equation}
as shown in Figure \ref{fig:CurvatureCorrection}(b). This is based on the consideration that the center of the tip is measured more accurately in z-coordinate, so it can be used as a reference for z correction. As examples, lattice planes in Figure \ref{fig:Summary}(c) and (e) become flat after curvature correction. The Sn composition profile after curvature correction is plotted in Figure \ref{fig:CurvatureCorrection}(c). The Sn composition is higher near the surface. This trend is fully consistent with the secondary-ion mass spectrometry (SIMS) results in the literature.\cite{dou2018investigation}

\subsection{Poisson-KNN method}

\begin{figure}[!htb]\centering
    \center{\includegraphics[width=\textwidth]
    {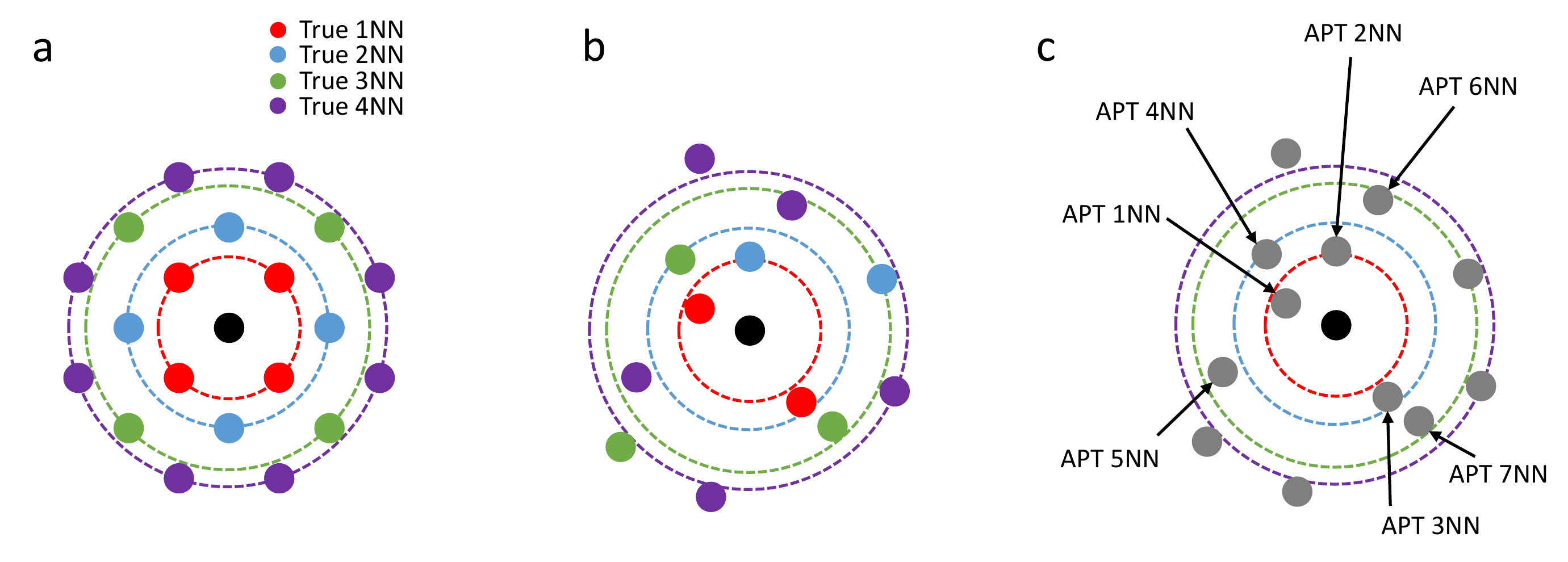}}
    \caption{A 2D schematic illustration of the true KNN vs. APT KNN. (a) An atomic map in a perfect crystal lattice with well-defined true KNN shells. (b) An atomic map acquired by APT with limited detection efficiency and spatial resolution. The true KNN can be missing or deviate from their perfect lattice sites. (c) Nominal KNN in APT data. Nominal KNN are ordered by the distance to the center reference atom.}
    \label{fig:APTKNN}
\end{figure}

Due to the low detection efficiency ($\sim$30\%) and relatively larger measurement error in atomic position for semiconductors, the nominal KNN in APT data do not fully represent the \emph{true} KNN of an atom in the ideal lattice, as schematically illustrated in \textbf{Figure \ref{fig:APTKNN}}. In a perfect crystal lattice, each atom typically has several true KNN in each shell. For example, a diamond cubic structure has 4 atoms in the 1NN shell, 12 in 2NN shell and 12 in 3NN shell.  Yet in the nominal KNN in APT analysis, each atom only has one 1NN, one 2NN, one 3NN etc., as ranked by the distance from this atom, because no two pairs of atoms have exactly the same distance due to experimental error in measuring atomic positions. 

In order to retrieve the \emph{true} KNN information from APT KNN, we propose a statistical method combining Poisson distribution and KNN analysis based on the crystal structure, which we call "Poisson-KNN method". In an ideal diamond cubic structure, each atom has 4 true 1NNs. If the detection efficiency is $\eta$, on average each atom should have $4\eta$ true 1NNs in the APT data. Assuming that the number of true 1NN atoms detected in APT follows Poisson distribution, the probability of detecting $k_1$ true 1NN is 
\begin{equation}
  P(k_{1},\lambda_{1})=\frac{\lambda_{1}^{k_{1}}e^{-\lambda_{1}}}{k_{1}!},
  \label{eqn:Poisson_1}
\end{equation}
where $\lambda_{1}=4\eta$ is the mean of the Poisson distribution. Here we chose Poisson statistics because measuring atoms from a KNN shell in APT translates to detecting those atoms within a certain time(of flight in z-axis)-space (in x-y plane) interval, analogous to the scenario of Poisson point process in a mathematical space. Furthermore, the discrete distribution of Poisson process matches the discrete number of detected KNN atoms. 

Now let us consider how to construct the true 1NN shell from the APT KNN data. We assume that, while the atomic positions are perturbed in the APT data, statistically the relative comparison of interatomic distances still holds true. That is, statistically APT KNN are indeed no closer to the reference atom than APT (K-1)NN in reality. Under this basic assumption, if an APT 1NN belongs to the true 1NN shell, a necessary and sufficient  condition is that we must have detected at least one true 1NN in the APT measurement. Such a probability is simply one minus the probability of detecting zero true 1NN, i.e. $[1-P(0,\lambda_{1})]$ based on Poisson statistics. Therefore, the probability for an APT 1NN to contribute to the true 1NN shell is $[1-P(0,\lambda_{1})]$. Similarly, if an APT 2NN contributes to the true 1NN shell, the APT 1NN must also belong to the true 1NN shell. Therefore, we must have detected at least two out of the four true 1NN in the APT measurement, and the corresponding probability is simply $[1-P(0,\lambda_{1})-P(1,\lambda_{1})]$. We can further extend this analysis to account for the contribution of APT KNN to the true 1NN shell,  and sum up their contributions to reconstruct the true 1NN shell:
\begin{equation}
\begin{split}
  True\;1NN = &\; [1-P(0,\lambda_{1})]\;of\;APT\;1NN\;+ \\ &\; [1-P(0,\lambda_{1})-P(1,\lambda_{1})]\;of\;APT\;2NN\;+ \\ &\; [1-P(0,\lambda_{1})-P(1,\lambda_{1})-P(2,\lambda_{1})]\;of\;APT\;3NN\;+ \; ... \\ &\;
 =\sum_{m=1}^{\infty} (P_{\geq}(m,\lambda_{1})\;of\;APT\;mNN),
  \label{eqn:True1NN_1}
\end{split}
\end{equation}
where the probability of detecting at least $m$ members of the true 1NN shell is
\begin{equation}
  P_{\geq}(m,\lambda_{1})=1- \sum_{k_{1}=0}^{m-1} P(k_{1},\lambda_{1}) ,
  \label{eqn:Poisson_1_at_least}
\end{equation}

Similarly, in an ideal diamond cubic structure, each atom has 12 true 2NNs, 12 true 3NNs, and 6 true 4NNs. Then on average each atom should have $12\eta$ true 2NNs, $12\eta$ true 3NNs and $6\eta$ true 4NNs in the APT data. Using a similar approach as evaluating the contribution of APT KNN to the true 1NN shell, the true 2NN, 3NN and 4NN shells can be reconstructed from APT KNN as
\begin{equation}
  True\;2NN = \sum_{m=1}^{\infty} (\sum_{k_{1}=0}^{m-1}P(k_{1},\lambda_{1})P_{\geq}(m-k_{1},\lambda_{2})\;of\;APT\;mNN),
  \label{eqn:True2NN_1}
\end{equation}
\begin{equation}
  True\;3NN = \sum_{m=1}^{\infty} (\sum_{k_{1}=0}^{m-1}\sum_{k_{2}=0}^{m-1-k_{1}}P(k_{1},\lambda_{1})P(k_{2},\lambda_{2})P_{\geq}(m-k_{1}-k_{2},\lambda_{3})\;of\;APT\;mNN),
  \label{eqn:True3NN_1}
\end{equation}
\begin{equation}
\begin{split}
  True\;4NN = &\sum_{m=1}^{\infty} (\sum_{k_{1}=0}^{m-1}\sum_{k_{2}=0}^{m-1-k_{1}}\sum_{k_{3}=0}^{m-1-k_{1}-k_{2}}\\
  &P(k_{1},\lambda_{1})P(k_{2},\lambda_{2})P(k_{3},\lambda_{3})P_{\geq}(m-k_{1}-k_{2}-k_{3},\lambda_{4})\;of\;APT\;mNN),
  \label{eqn:True4NN_1}
\end{split}
\end{equation}
where $\lambda_{2}=\lambda_{3}=12\eta$, $\lambda_{4}=6\eta$.

\begin{figure}[!htb]\centering
    \center{\includegraphics[width=\textwidth]
    {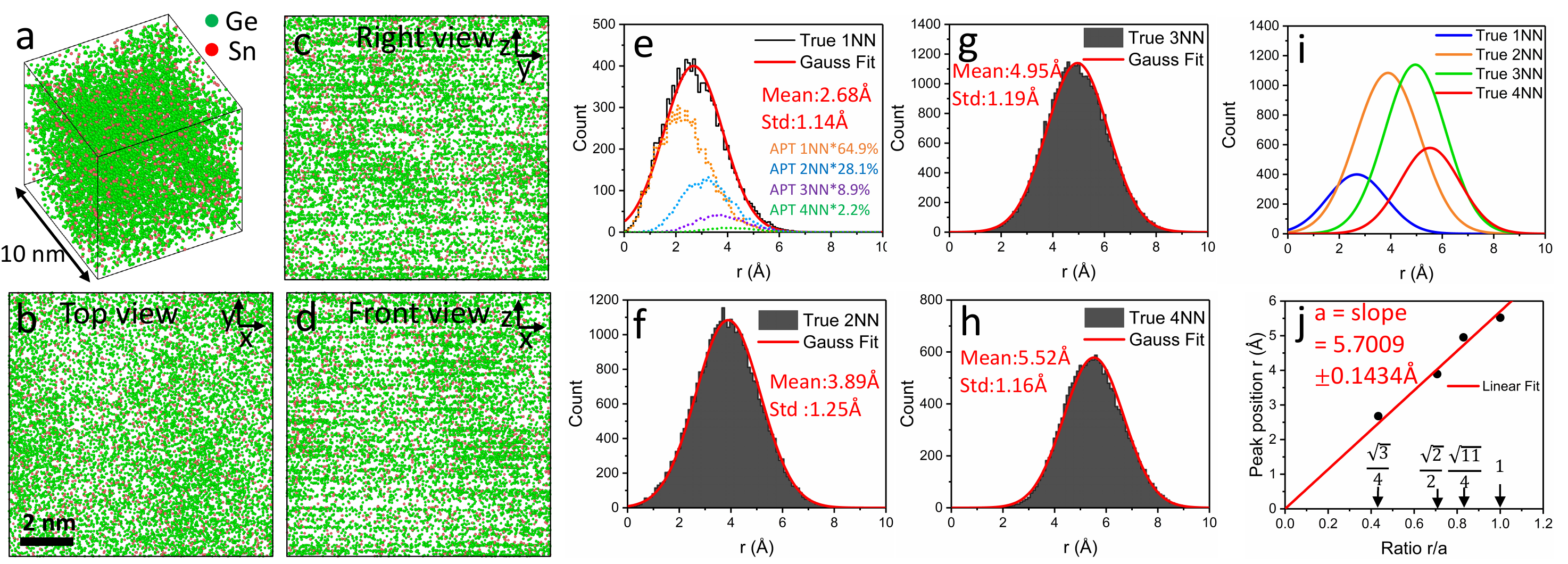}}
    \caption{(a) Curvature corrected atomic map of a 10 nm$\times$10 nm$\times$10 nm nanocube around z=200 nm. (b)-(d) Top view, right view and front view of (a). (e)-(h) True 1NN-4NN histograms retrieved by Possion-KNN method from (a). (e) also shows the constituent APT 1NN-4NN in constructing the true 1NN histogram. (i) Gaussian fit of true 1NN-4NN distribution of atoms in (e)-(h). (j) Lattice constant fitting using peak positions in (i).}
    \label{fig:PoissonKNN}
\end{figure}

\begin{table}
  \caption{True KNN shells constructed from statistically weighted combination of APT KNN for the regions shown in Figure \ref{fig:PoissonKNN}}
  \label{tbl:PoissonKNN_Component}
  \includegraphics[width=\linewidth]
  {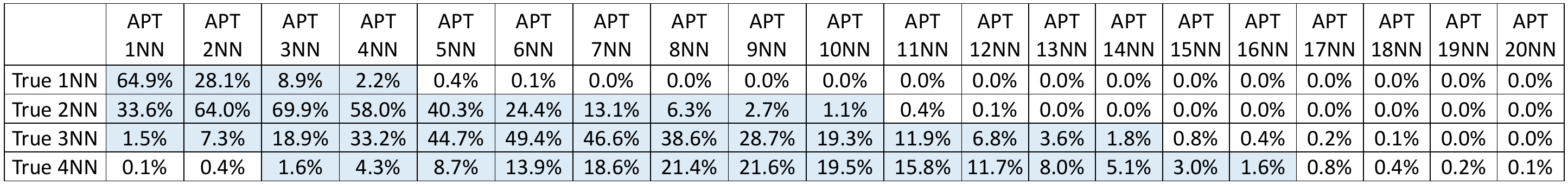}
\end{table}

A major difference from true 1NN shell construction is that now APT mNN may contribute to the true KNN shell even if m is smaller than K. For example, an APT 1NN might actually belong to the true 2NN shell. For this to happen, we must have (a) detected 0 true 1NN in the APT measurement (since even the APT 1NN belongs to the true 2NN shell); (b) detected at least one of the true 2NNs. Therefore, Equation (5) for constructing the true 2NN includes the product of two probability terms. Similarly, Equations (6) and (7) include the products of 3 and 4 probability terms, respectively, for constructing true 3NN and 4NN shells.

Therefore, using APT KNN histograms derived from the raw APT data, we can apply Equations (3)-(7) to reconstruct the true 1NN-4NN histograms. From the peak positions of the true 1NN-4NN histograms, we can obtain the corresponding KNN shell radii and do a linear fit to derive the local lattice constant based on the proportionality of the former to the latter. The derived lattice constant of a 10x10x10 $nm^3$ also provides a semi-quantitative measurement of the corresponding local strain when compared to the ideally relaxed lattice constant. 

We first test this method on the APT data of epitaxial Ge on Si grown by a typical two-step growth method, which has 0.2\% in-plane tensile strain measured by X-ray diffraction.\cite{liu2004deformation} The derived lattice constant of 5.6910$\pm$0.1634 \AA \; using the aforementioned statistical true KNN analysis of APT data is in good agreement with that of 0.2\% tensile strained Ge (5.6692 \AA). This agreement validates our statistical method in reconstructing the true KNN shells (see Supporting Information). 

We then apply this true KNN retrieval approach to the GeSn alloy. \textbf{Figure \ref{fig:PoissonKNN}}(a)-(d) are the atomic maps of a 10x10x10 $nm^3$ nanocube after curvature correction. The size of the nanocube was chosen to include a large enough number of atoms (>10,000) for statistical validity. This factor also sets the spatial resolution limit of our approach for SRO mapping since overly small regions would lose the validity of statistical methods. In this nanocube, the Sn composition is $x_{Sn} = 14.4\%$, and the measured atomic density is $\rho=1.08\times10^{22}/cm^3$. By comparison, ideally the lattice constant of GeSn with such Sn composition should be $a_{ideal}=5.7777$ \AA \;by
\begin{equation}
  a_{ideal} = a_{Ge}\times(1-x_{Sn})+a_{Sn}\times x_{Sn},
  \label{eqn:IdealGeSnLatticeConstant}
\end{equation}
where Ge lattice constant $a_{Ge}=5.6579$ \AA\  \cite{madelung2012semiconductors} and \textalpha-Sn lattice constant $a_{Sn}=6.4892$ \AA,\cite{madelung2012semiconductors} both at room temperature. Here Vegard's law is utilized because recent literature has indicated that the bowing parameter of GeSn is practically 0 within the experimental error range.\cite{xu2017deviations} Even though APT was conducted at low temperature (40 K), the actual temperature upon atomic ablation is estimated to be 300-400 K. Furthermore, for both Ge and \textalpha-Sn, the difference between lattice constant at low and room temperature is less then 0.2\%, within the error bar of lattice constant fitting using APT data. Therefore, the temperature effect on lattice constant is ignored in this paper. The ideal atomic density for this Sn composition would be $\rho_{ideal}=4.15\times10^{22}/cm^3$ by
\begin{equation}
  \rho_{ideal} = \frac{8}{a_{ideal}^3}.
  \label{eqn:IdealGeSnAtomicDensity}
\end{equation}
Correspondingly, the detection efficiency is calculated to be $\eta=26.16\%$ by
\begin{equation}
  \eta = \frac{\rho}{\rho_{ideal}}.
  \label{eqn:DetectionEfficiency}
\end{equation}
\textbf{Table \ref{tbl:PoissonKNN_Component}} tabulates the true 1NN-4NN of this nanocube constructed from statistically weighted combination of APT KNN by Equation (\ref{eqn:True1NN_1})-(\ref{eqn:True4NN_1}). Only those APT KNN with a weight larger than 1\% are used. These series converge quickly, so using APT 1NN up to 20 NN is well sufficient for TRUE 1NN-4NN analyses. The histograms of the true 1NN-4NN are plotted in Figure \ref{fig:PoissonKNN}(e)-(h). All of them subject very well to Gaussian distribution. The standard deviations of true 1NN-4NN distributions are $\sim$1.2 \AA, slightly smaller than the difference in radii between 1NN and 2NN shells but larger than those between 2NN-4NN shells. Therefore, the true 1NN-4NN distributions are overlapping with each other as shown in Figure \ref{fig:PoissonKNN}(i). This is the reason why we cannot distinguish true 1NN-4NN peaks directly from RDF analysis of the APT data. The relatively crude methods to distinguish KNN shells, such as artificially cutting a line midway between the theoretical radii of the KNN and (K+1)NN shells to accommodate the perturbed atomic positions,\cite{ceguerra2012short} are not applicable in our case, either, due to the artificial truncation of the relatively broad 1NN-4NN Gaussian peaks. 

For a diamond cubic structure with a lattice constant $a$, the radii of 1NN-4NN shells are $\frac{\sqrt{3}}{4}a$, $\frac{\sqrt{2}}{2}a$, $\frac{\sqrt{11}}{4}a$, and $a$, respectively. By doing a linear fitting (with zero intercept) of the peak positions of true 1NN-4NN histograms, we can derive the lattice constant $a = 5.7009\pm0.1434$ \AA \;with 95\% confidence (see Figure \ref{fig:PoissonKNN}(j)). The nominal strain $\varepsilon=-0.013\pm0.025$, given by
\begin{equation}
  \varepsilon = \frac{a-a_{ideal}}{a_{ideal}}.
  \label{eqn:nominal strain}
\end{equation}
The derived compressive strain of $\sim$ -1\% from APT is again largely consistent with the X-ray diffraction results of -0.6\%,\cite{dou2018investigation} although the error bar is relatively large. Therefore, the strain derived using this approach can still semi-quantitatively represent the local strain at nanoscale, offering strain mapping capability in addition to SRO mapping. 

\subsection{Short-range order}
After retrieving true KNN information, we can examine the SRO distribution in different nanocubes of the sample by comparing the Sn composition in Sn atoms' true KNN shells with that of the same nanocube. As mentioned earlier, for GeSn alloys we define SRO parameter as
\begin{equation}
  \alpha^{KNN}_{Sn-Sn} = \frac{P^{KNN}_{Sn-Sn}}{x_{Sn}}
    \begin{cases}
      >1, & Sn \;atoms \; preferred \;in \;another \;Sn \;atom's \;KNN \;shell \\
      =1, & random \;alloy:\;no \; preferred \; atomic \; species \; in \;the \;KNN \;shell \\
      <1, & Sn \;atoms \;unpreferred  \;in \;another \;Sn \;atom's \;KNN \;shell 
    \end{cases},
  \label{eqn:SRO_Parameter_KNN}
\end{equation}
 where $P^{KNN}_{Sn-Sn}$ is the probability for a Sn atom to have another Sn atom as a true KNN, and $x_{Sn}$ is the Sn at.\% in the same nanocube region. For example, $\alpha^{KNN}_{Sn-Sn}$>1 indicates more preference for a Sn atom to have another Sn atom in the KNN shell than a random alloy. Especially, $\alpha^{1NN}_{Sn-Sn}$>1 indicates a tendency of Sn-Sn clustering. 
 
 As a reference, Random Labelling method \cite {gault2012atom,ceguerra2010three,marceau2010evolution} can be used to generate a true random alloy to test our Poisson-KNN method for SRO analyses. We therefore randomly shuffle the position of Ge and Sn atoms in nanocube A in \textbf{Figure \ref{fig:Summary}}(f) for 100 times, and calculate the SRO parameters after each shuffle. The average SRO parameters of these 100 configurations are indeed almost 1:  $\alpha^{1NN}_{Sn-Sn,avg}$=0.999, $\alpha^{2NN}_{Sn-Sn,avg}$=0.999, $\alpha^{3NN}_{Sn-Sn,avg}$=1.000, $\alpha^{4NN}_{Sn-Sn,avg}$=0.998, which is expected because the average of 100 randomly shuffled atom configurations should be a true random alloy. This result verifies the baseline of the Sn-Sn KNN SRO parameters derived from our Poisson-KNN method.
 
\begin{figure}[!htb]\centering
    \center{\includegraphics[scale=0.7]
    {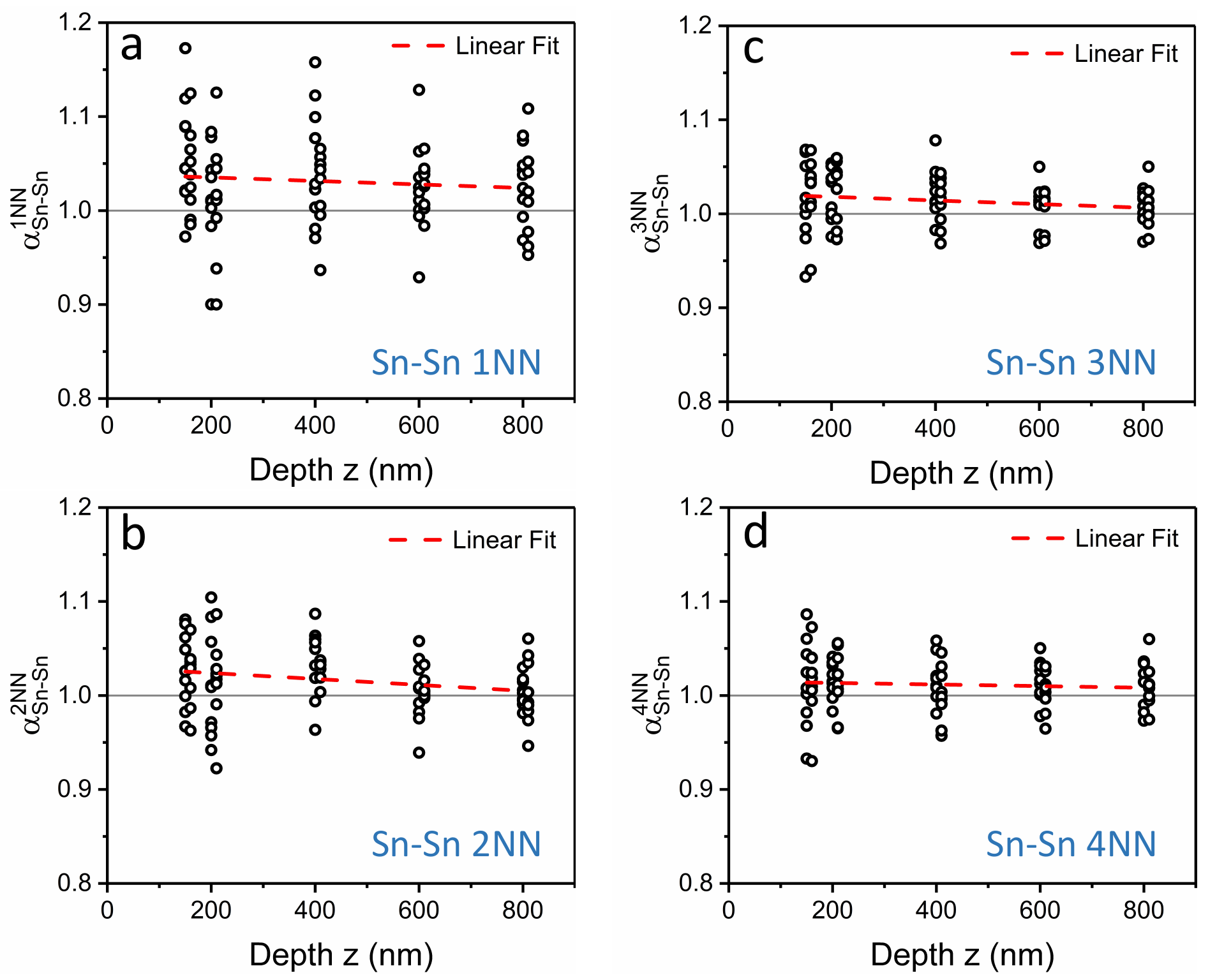}}
    \caption{(a)-(d) $\alpha^{1NN}_{Sn-Sn}$, $\alpha^{2NN}_{Sn-Sn}$, $\alpha^{3NN}_{Sn-Sn}$, $\alpha^{4NN}_{Sn-Sn}$ versus depth. Ninety 10 nm$\times$10 nm$\times$10 nm nanocubes are examined after curvature correction.}
    \label{fig:SROMapCurvatureCorrected}
\end{figure}

 Ninety 10$\times$10$\times$10 $nm^3$ nanocubes at different depths of the GeSn film are then studied and their SRO parameters are partially mapped in Figure 1(f) as well as summarized in \textbf{Figure \ref{fig:SROMapCurvatureCorrected}}. The 3D SRO mapping in Figure 1(f) at 195-215 nm and 595-615 nm depths shows a relatively large fluctuation of Sn-Sn 1NN SRO parameter $\alpha^{KNN}_{Sn-Sn}$ from 0.85 to 1.15 even at the same depth, indicating the non-uniform SRO at nanoscale. Sometimes even adjacent nanocubes can have drastically different $\alpha^{KNN}_{Sn-Sn}$, e.g. the dark red nanocube A and the adjacent dark blue nanocube on its right. This Sn-Sn 1NN SRO fluctuation in GeSn, together with the theoretical prediction of large band structure variation with SRO,\cite{cao2020short} contributes to the bandedge softening compared to pure Ge, as commonly observed in GeSn absorption spectra as well as  PL peak width$>>k_BT$ thermal broadening even at 10 K.\cite{olorunsola2021impact,zhou2020electrically,ye2015absorption,chen2014structural} The data summarized in Figure \ref{fig:SROMapCurvatureCorrected} further shows that Sn-Sn 1NN SRO parameter $\alpha^{1NN}_{Sn-Sn}$ decreases from 1.052$\pm$0.051 to 1.026$\pm$0.045 as the depth increase from 150 to 800 nm under the surface, with 18 10 nm $\times$10 nm$\times$10 nm blocks examined at each depth. These results indicate more preference towards Sn-Sn clustering near the surface, consistent with the tendency of Sn segregation on the surface. Furthermore, Sn-Sn 1NN-4NN are all slightly more favored than those in random alloys throughout the depth of the GeSn film, and such preference weakens and becomes closer to that in random alloys as the shell number K increases from 1 to 4. Such a slight preference for a Sn atom to have other Sn atoms in 1NN-4NN shells is compensated by a slight less preference of having Ge atoms the same shells. The preference of Sn-Sn 2NN over Sn-Ge 2NN from the Poisson-KNN analysis also qualitatively agrees with previous EXAFS results in Ref.\cite{gencarelli2015extended}  .
 
\begin{figure}[!htb]\centering
    \center{\includegraphics[width=\textwidth]
    {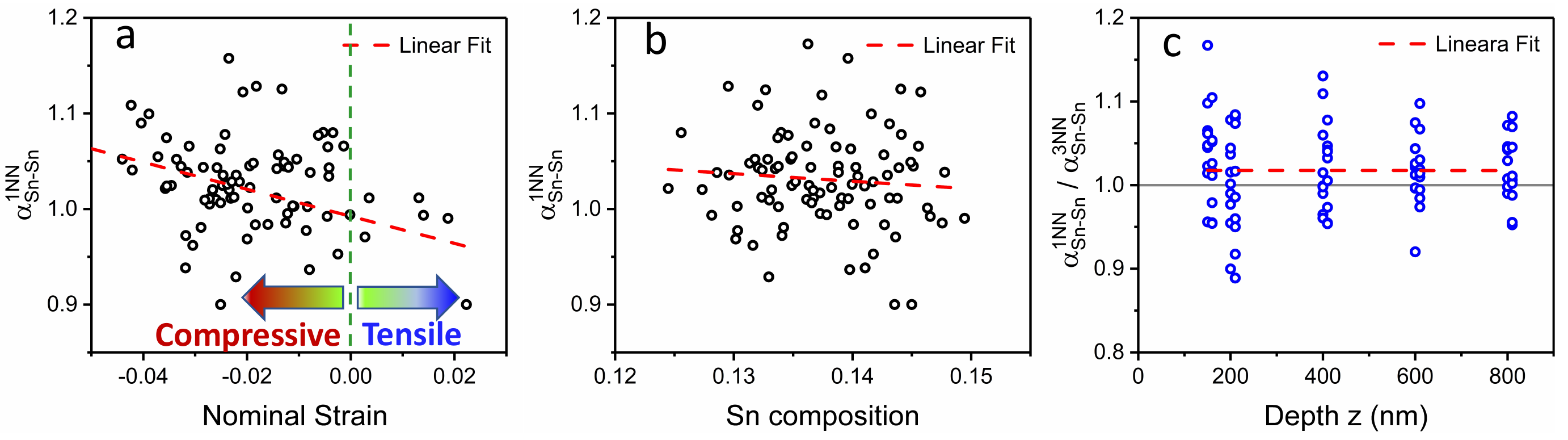}}
    \caption{(a) $\alpha^{1NN}_{Sn-Sn}$ versus strain. (b) $\alpha^{1NN}_{Sn-Sn}$ versus Sn composition. (c) $\alpha^{1NN}_{Sn-Sn}/\alpha^{3NN}_{Sn-Sn}$ versus depth. Ninety 10 nm$\times$10 nm$\times$10 nm nanocubes are examined after curvature correction. 10\% of data are outliers and are not included in (a).}
    \label{fig:SROMapCurvatureCorrected_2}
\end{figure}

 We then investigate the dependence of Sn-Sn 1NN SRO parameter on local strain and Sn composition. $\alpha^{1NN}_{Sn-Sn}$ versus nominal strain is plotted in \textbf{Figure \ref{fig:SROMapCurvatureCorrected_2}}(a), showing that compressive strain tends to favor Sn-Sn clustering (larger $\alpha^{1NN}_{Sn-Sn}$), while strain relaxation or tensile strain favors declustering of Sn-Sn 1NN. Most of the data points in this figure correspond to compressive strain, consistent with the XRD measurement of an -0.6 \% compressive strain in the GeSn film. More detailed nano-scale strain mapping at different depths are shown in the Supporting Information, where the average compressive strain of -2\% is somewhat overestimated compared to the XRD result of -0.6\%. Further calibration with the average XRD strain data may provide more accurate strain mapping from Poisson-KNN analyses of the APT data. We also note that while the trend of Sn-Sn 1NN SRO vs. local strain is clear in \textbf{Figure \ref{fig:SROMapCurvatureCorrected_2}}(a), this numerical relation is semi-quantitative because of the error bars of the derived strain from APT data as shown previously. Seven out of the 90 strain data points are outliers due to relatively large errors and therefore are not included in Figure \ref{fig:SROMapCurvatureCorrected_2}(a).  Interestingly, in contrast to the strain-dependence, the Sn-Sn 1NN SRO parameter hardly depends on the local Sn composition (see Figure \ref{fig:SROMapCurvatureCorrected_2}(b)). This result indicates that local strain plays a more significant role in determining the tendency of Sn-Sn clustering/declustering than the composition fluctuations between 12.5 and 15 \% Sn. 
 
 An important theoretical prediction of GeSn at thermodynamic equilibrium is that random alloys would have a preference of Sn-Sn 1NN over Sn-Sn 3NN, while more stable SRO structures would behave in an opposite way with Sn-Sn 3NN strongly preferred over Sn-Sn 1NN.\cite{cao2020short} Therefore, we plot the ratio of $\alpha^{1NN}_{Sn-Sn}$ to $\alpha^{3NN}_{Sn-Sn}$ in Figure \ref{fig:SROMapCurvatureCorrected_2}(c) to compare with the theoretical predictions. This $\alpha^{1NN}_{Sn-Sn}$ to $\alpha^{3NN}_{Sn- Sn}$ ratio actually fluctuates around 1 and does not show strong preference of either 1NN or 3NN , indicating that in reality the GeSn film is midway between a truly random alloy and a SRO structure at thermodynamic equilibrium (with minimized energy). This result is  reasonable since CVD growth is neither a thermodynamic equilibrium process nor a totally random process. 
 
\subsection{Rearrangement of atoms to diamond cubic lattice sites}

We also further validate this Poisson-KNN statistical approach by fitting and rearranging the atomic positions to the diamond cubic lattice. Methods such as Fourier transform \cite{camus1995method,kumar2013atomic,vurpillot2001structural} and spatial distribution map \cite{moody2009qualification,moody2011lattice,devaraj2018three,moody2014atomically} have been demonstrated to be able to rearrange atoms to perfect lattice in metallic alloys, but those methods are not applicable to semiconductors due to the larger spatial error. We therefore adopted an algorithm based on least square fit to rearrange atoms to nearby perfect lattice sites, which show a very good consistency in Sn-Sn SRO analyses with the aforementioned Poisson-KNN method before atomic rearrangement. The procedure is described in detail in this section.

\begin{figure}[!htb]\centering
    \center{\includegraphics[scale=0.60]
    {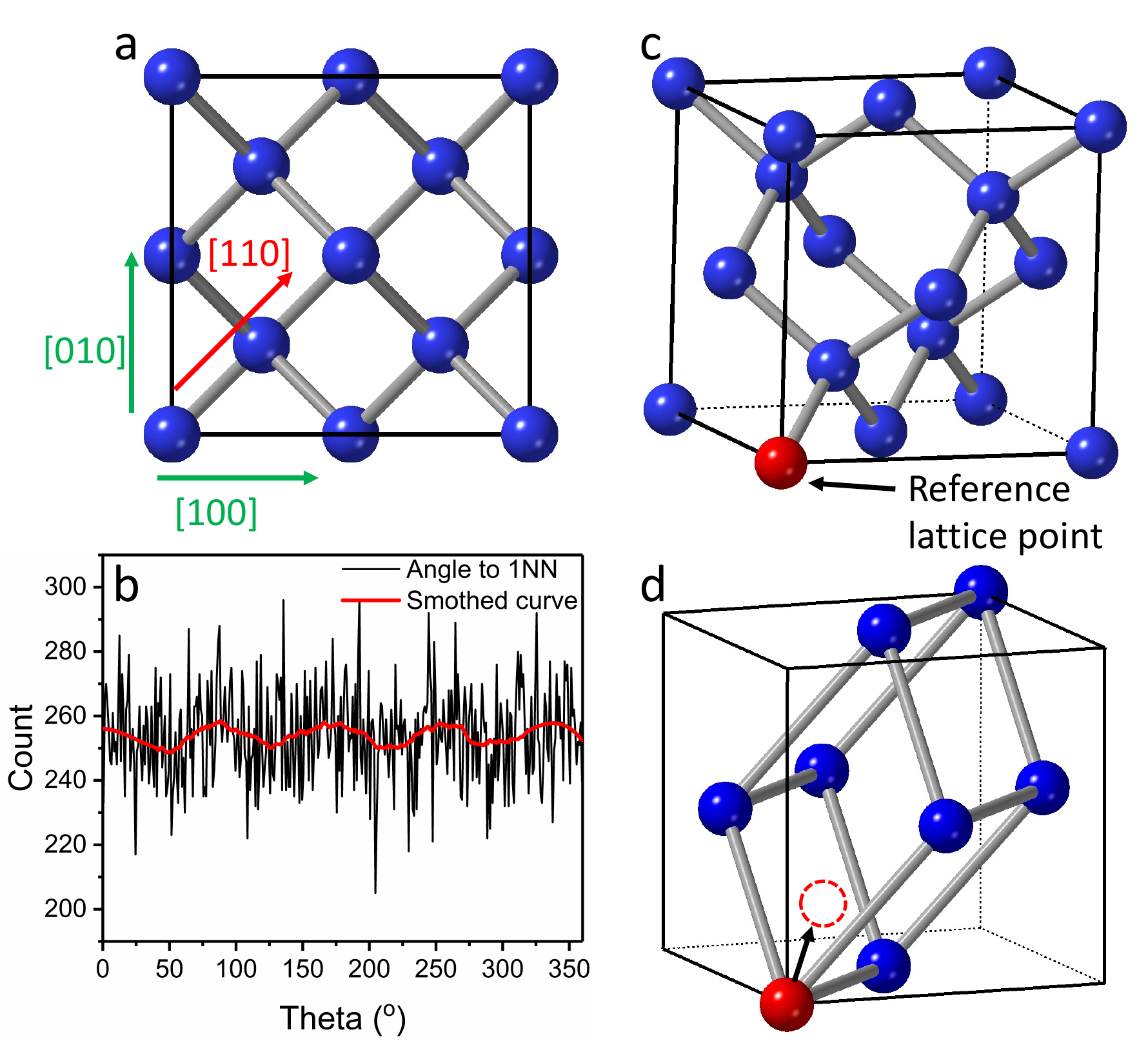}}
    \caption{(a) Diamond cubic structure projected onto (001) plane. (b) Statistical distribution of angles to 1NN in reference to the x-axis of the original APT data. (c) A unit cell of diamond cubic structure. (d) A primitive cell of diamond cubic structure.}
    \label{fig:Rotation}
\end{figure}

To fit the atomic sites in single crystal diamond cubic lattice, we will first determine the crystallographic orientation in the x-y plane, considering that the z-axis is along [001] growth direction of the GeSn film. After projecting the atoms in a diamond cubic structure onto (001) plane, the close-packed direction is <110> as shown in \textbf{Figure \ref{fig:Rotation}}(a). i.e. the direction with the shortest distance between atoms. Therefore, the <110> orientations can be determined by making a histogram of the directions from an atom to its APT 1NN projected on the x-y plane. The peaks of the histogram would correspond to the <110> directions. In order to do so, we studied a volume with 20 nm$\times$20 nm$\times$20 nm dimension. All atoms in this volume were projected onto x-y plane ((001) plane), and the direction from each atom to its 1NN were recorded and converted to the angle in reference to x axis. Figure \ref{fig:Rotation}(b) shows the histogram of the angle to 1NN. A 3rd order Savitzky–Golay filter was used to smooth the curve. The smoothed curve has clear period of $\sim$90\degree, and the peaks are around 90\degree, 180\degree, 270\degree and 360\degree. This means the x axis is in [110] direction. In order to transform x axis to the [100] direction, we should rotate the sample by 45\degree around the z axis.

\begin{figure}[!htb]\centering
    \center{\includegraphics[scale=0.70]
    {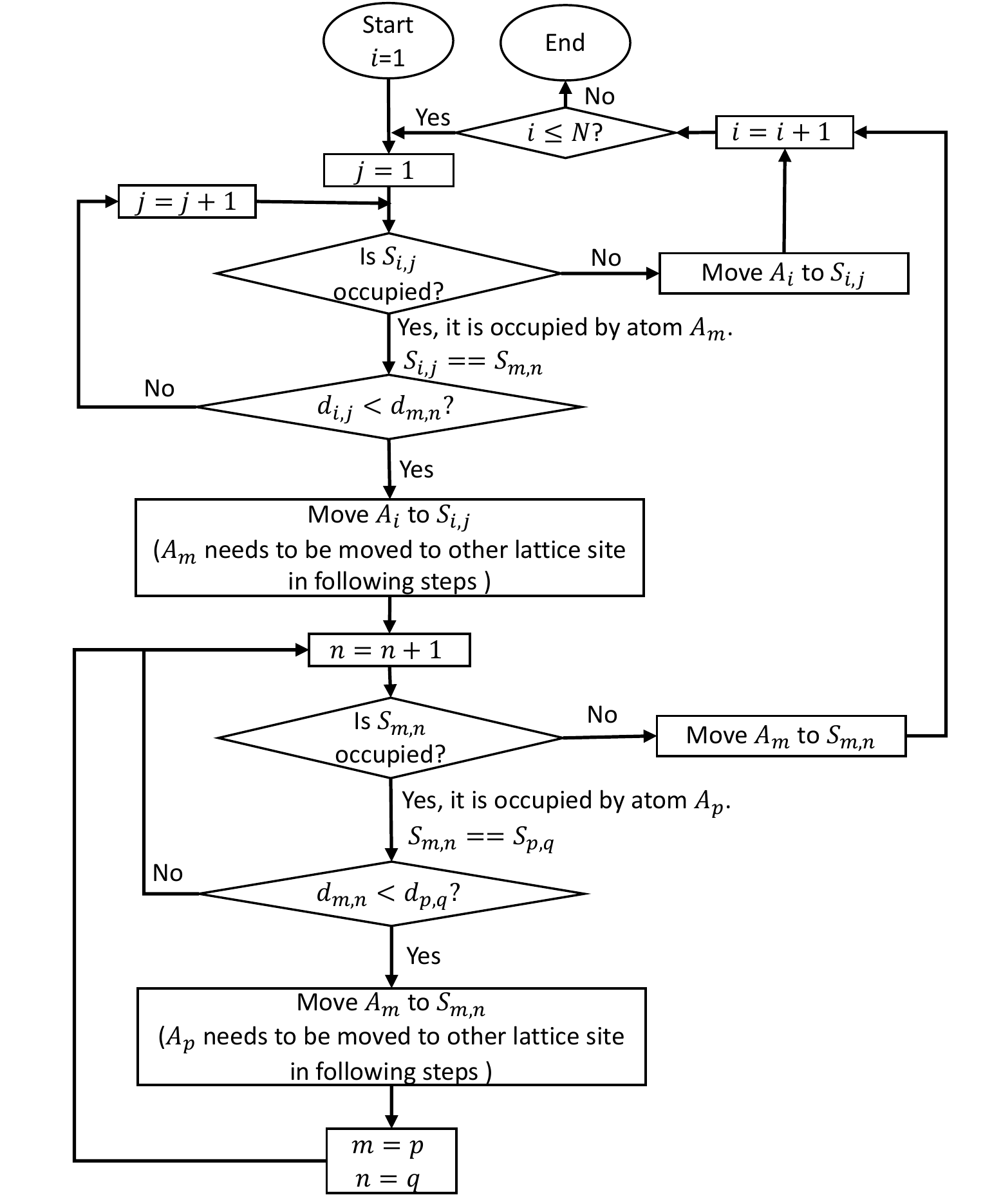}}
    \caption{Algorithm to rearrange atoms to nearby perfect diamond cubic lattice sites. $A_i$: the $i^{th}$ atom. $S_{i,j}$: the $j^{th}$ narest lattice site of atom $A_i$. $d_{i,j}$: the distance from the original position of atom $A_i$ to lattice site $S_{i,j}$. $N$: total number of atoms.}
    \label{fig:RearrangeAlgorithm}
\end{figure}

Given the  [100], [010] and [001] directions determined above and the lattice constant derived from Equation (\ref{eqn:IdealGeSnLatticeConstant}), the last piece of information required to determine a diamond cubic crystal lattice is the position of the reference lattice point. Once the position of the reference lattice point is known, all the other lattice sites of single crystalline epitaxial GeSn is determined as shown in Figure \ref{fig:Rotation}(c). And once the crystal lattice is determined, the atoms in the APT data can be rearranged to nearby lattice sites. So we can scan the position of the reference lattice point and select the one that gives minimum root mean square (RMS) of the rearrangement distance to match the atoms to the lattice points:
\begin{equation}
  min(\sqrt{\frac{\sum_{i=1}^{N}(r_{i}-r'_{i})^2}{N}}),
  \label{eqn:RootMeanSqureDisplacement}
\end{equation}
where $r_i$ is the original position of an atom, $r'_i$ is the rearranged position of that atom on the lattice site, and $N$ is the total number of atoms.
Considering the repetition of the crystal lattice, the position of the reference lattice point only needs to be swept within a volume defined by primitive cell as highlighted in Figure \ref{fig:Rotation}(d). The algorithm to rearrange atoms to a known perfect lattice is shown in \textbf{Figure \ref{fig:RearrangeAlgorithm}}. It is preferred to rearrange atom to its 1st nearest lattice site, but if multiple atoms are moved to same lattice site, only the atom with smallest displacement occupies this lattice site. Correspondingly, the other atoms will be moved to their 2nd or further nearest lattice site. 

\begin{figure}[!htb]\centering
    \center{\includegraphics[width=\textwidth]
    {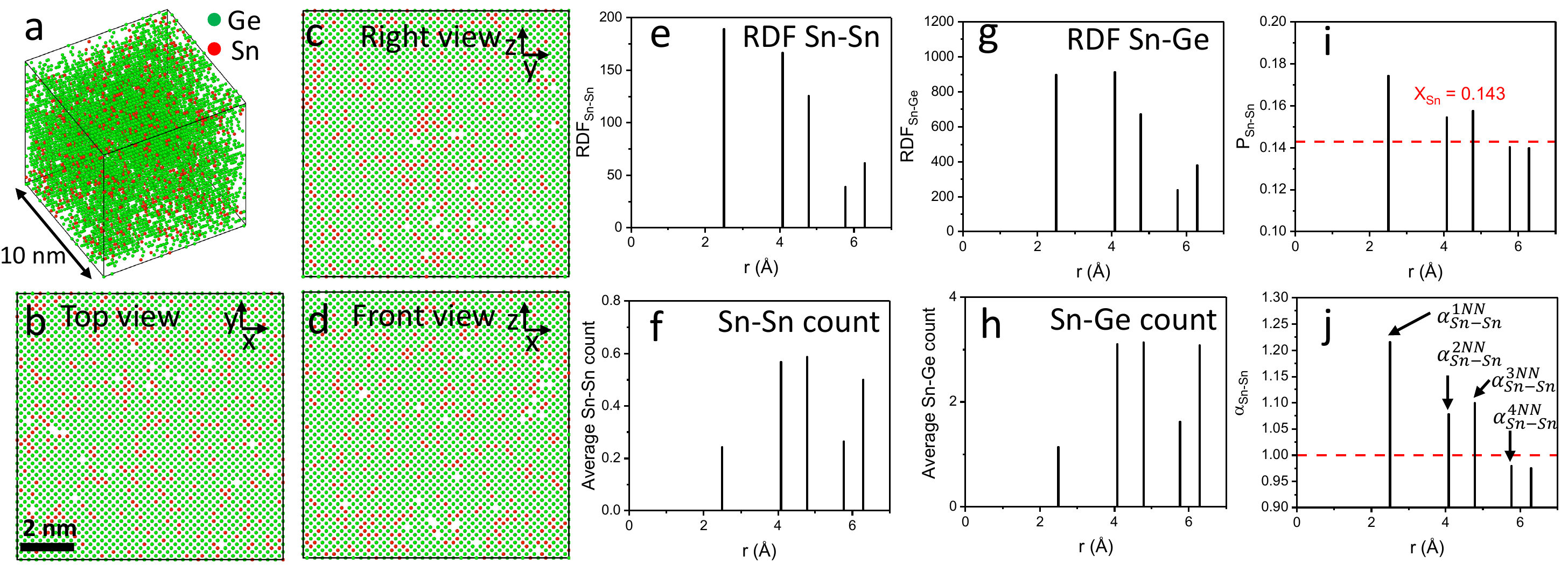}}
    \caption{(a) Atomic map of a 10 nm×10 nm×10 nm nanocube around z=200 nm after rearrangement to diamond cubic lattice sites. (b)-(d) Top view, right view and front view of (a). (e), (g) Radial distribution function (RDF) of Sn-Sn and Sn-Ge pair in (a). (f), (h) Distribution of average count of Sn-Sn and Sn-Ge pairs. (i) Probability for a Sn atom to have another Sn atom in its KNN shells. (j) Short range order parameters for Sn-Sn KNN.}
    \label{fig:Rearranged_1}
\end{figure}

\textbf{Figures \ref{fig:Rearranged_1}}(a)-(d) show the atomic map of a 10 nm×10 nm×10 nm nanocube (nanocube A as shown in Figure \ref{fig:Summary}) around z=200 nm after rearrangement to diamond cubic lattice sites. The RMS of rearrangement distance is 1.61 \AA. We note that this is comparable to the Gaussian standard deviation in fitting the true KNN histograms in Figure \ref{fig:PoissonKNN}. Figures \ref{fig:Rearranged_1}(e)-(h) show the RDF and the distribution of average counts of Sn-Sn and Sn-Ge pairs in 1NN to 4NN shells. The probability for a Sn atom to have another Sn atom in the KNN shell (shown in Figure \ref{fig:Rearranged_1}(i)) is defined as
\begin{equation}
\begin{split}
  P^{KNN}_{Sn-Sn} &= \frac{RDF^{KNN}_{Sn-Sn}}{RDF^{KNN}_{Sn-Sn}+RDF^{KNN}_{Sn-Ge}}\\
  &=\frac{Average\;Sn-Sn\;count}{(Average\;Sn-Sn\;count)\;+\;(Average\;Sn-Ge\;count)}\;in\;KNN\;shell,
  \label{eqn:Sn-Sn_Probability}
\end{split}
\end{equation}
and the SRO parameter (shown in Figure  \ref{fig:Rearranged_1}(j)) is defined the same way as earlier text:
\begin{equation}
  \alpha^{KNN}_{Sn-Sn} = \frac{P^{KNN}_{Sn-Sn}}{x_{Sn}}.
  \label{eqn:SRO_Parameter}
\end{equation}

Again, as a reference, Random labelling method \cite{gault2012atom,ceguerra2010three,marceau2010evolution} can be used to generate a true random alloy to test our lattice fitting method for SRO analyses. We randomly shuffle the position of Ge and Sn atoms in this nanocube for 100 times, and the average $\alpha_{Sn-Sn}$ indeed becomes 1 in the corresponding RDFs (see Supporting Information), which is expected because the average of 100 randomly shuffled atom configurations should be a true random alloy. This verifies our atomic position fitting algorithm to calculate Sn-Sn KNN SRO parameters. 

\begin{figure}[!htb]\centering
    \center{\includegraphics[width=\textwidth]
    {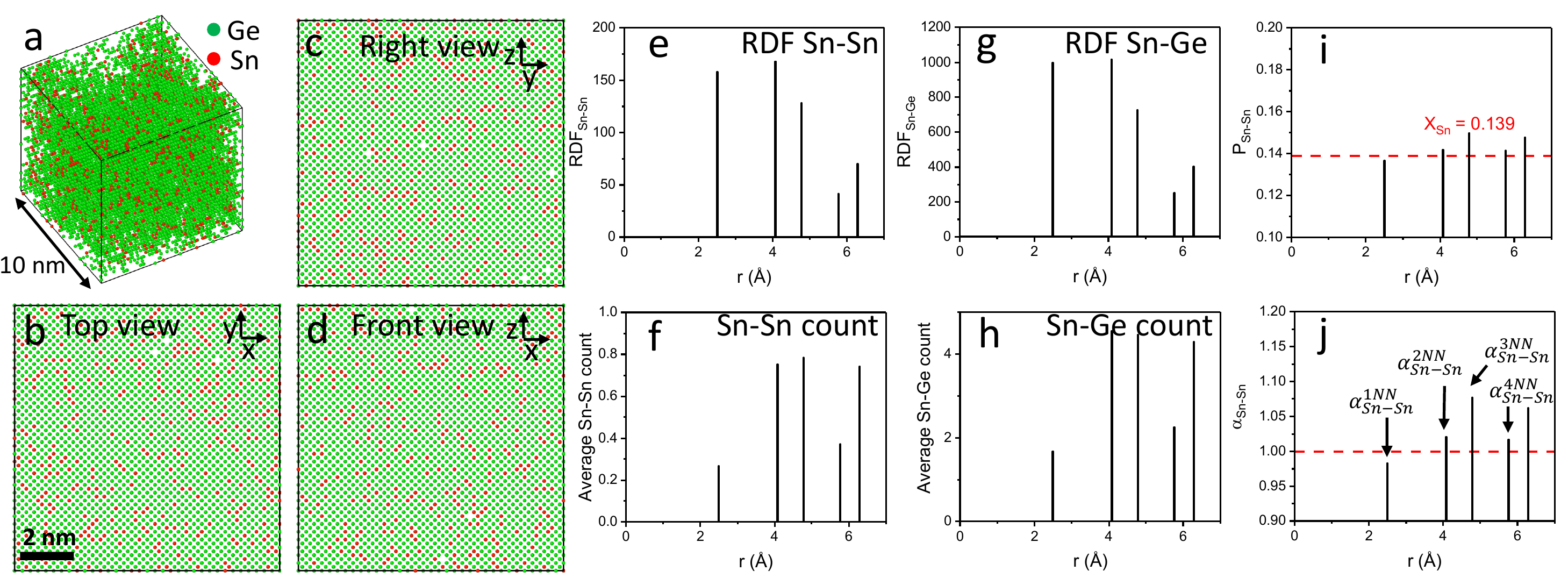}}
    \caption{(a) Atomic map of a 10 nm×10 nm×10 nm nanocube around z=600 nm after rearrangement to diamond cubic lattice sites. (b)-(d) Top view, right view and front view of (a). (e), (g) Radial distribution function (RDF) of Sn-Sn and Sn-Ge pair in (a). (f), (h) Distribution of average count of Sn-Sn and Sn-Ge pair. (i) Probability for a Sn atom to have another Sn atom in its KNN shells. (j) Short range order parameter for Sn-Sn KNN.}
    \label{fig:Rearranged_2}
\end{figure}

Based on this lattice fitting approach, the first 4 peaks in Figure \ref{fig:Rearranged_1}(j) are $\alpha^{1NN}_{Sn-Sn}$, $\alpha^{2NN}_{Sn-Sn}$, $\alpha^{3NN}_{Sn-Sn}$ and $\alpha^{4NN}_{Sn-Sn}$. The $\alpha^{1NN}_{Sn-Sn}$ is much higher than 1 (labelled as the red dot line), indicating a strong Sn-Sn clustering in this region, and this is also consistent with the $\alpha^{1NN}_{Sn-Sn}$ calculated using Poisson-KNN method before rearrangement of the atoms (see nanocube A in Figure \ref{fig:Summary}(f)).  Another nanocube around z=600 nm (nanocube B as shown in Figure \ref{fig:Summary}) is examined and shown in Figure \textbf{\ref{fig:Rearranged_2}}. The RMS of rearrangement distance is 1.67 \AA, similar to the nanocube at $\sim$200 nm depth. The low $\alpha^{1NN}_{Sn-Sn}$ peak and high $\alpha^{3NN}_{Sn-Sn}$ peak indicates the tendency of Sn-Sn repulsion/declustering in this region. Clearly, the atomic position fitting approach also yields a notable fluctuation of Sn-Sn 1NN SRO parameters, confirming the results from our Poisson-KNN SRO mapping approach.

\begin{figure}[!htb]\centering
    \center{\includegraphics[scale=0.60]
    {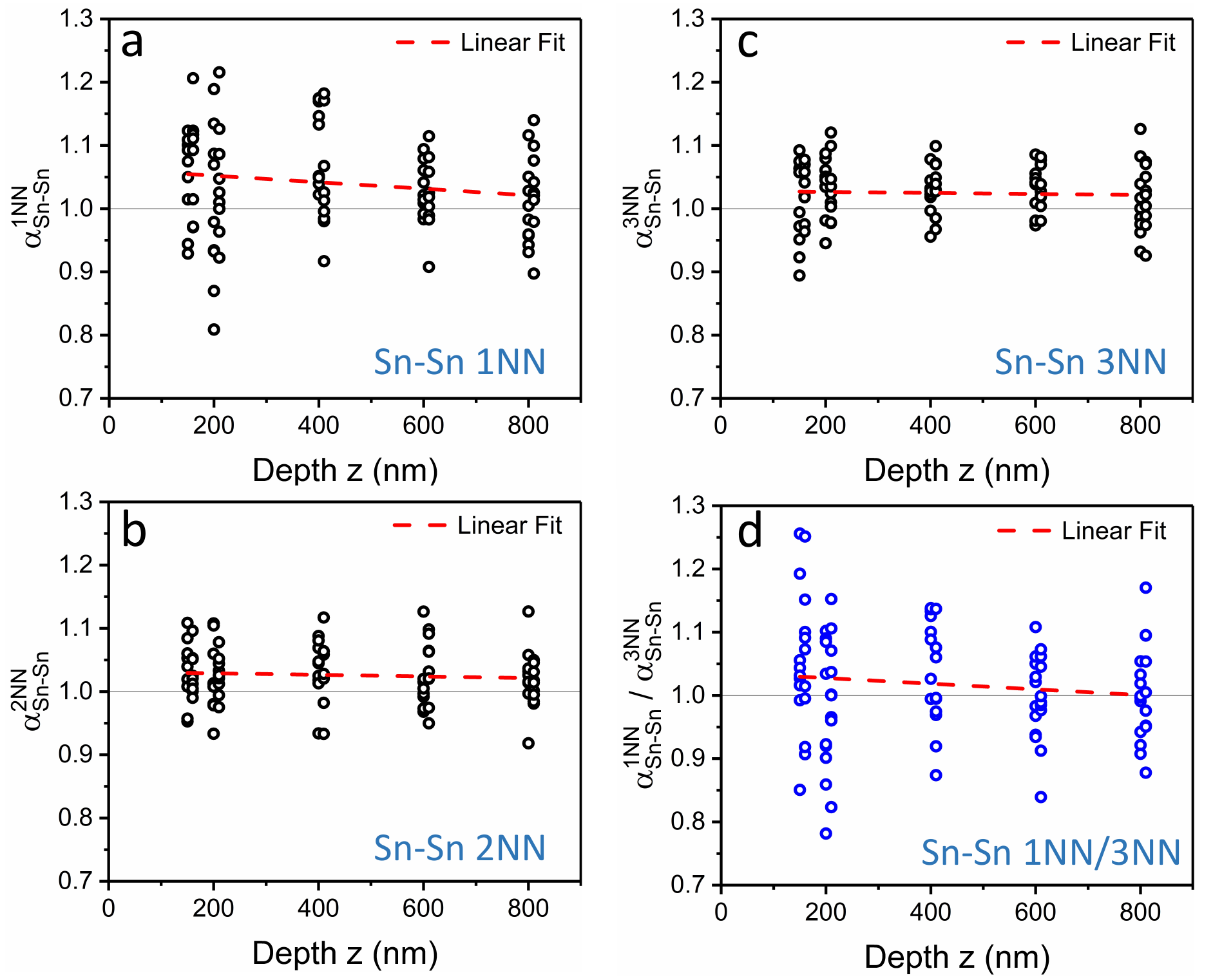}}
    \caption{(a)-(e) $\alpha^{1NN}_{Sn-Sn}$, $\alpha^{2NN}_{Sn-Sn}$, $\alpha^{3NN}_{Sn-Sn}$, and $\alpha^{1NN}_{Sn-Sn}/\alpha^{3NN}_{Sn-Sn}$ versus depth. Ninety 10 nm×10 nm×10 nm blocks are examined after after rearrangement to diamond cubic lattice sites.}
    \label{fig:SROMapRearranged}
\end{figure}

The same 90 10 nm$\times$10 nm$\times$10 nm nanocubes as used earlier for Poisson-KNN analyses are also subjected to atomic position fitting. Their SRO parameter after lattice fitting and atomic rearrangmenet are summarized in \textbf{Figure \ref{fig:SROMapRearranged}}. $\alpha^{1NN}_{Sn-Sn}$ decreases from 1.065$\pm$0.075 to 1.015$\pm$0.066 as the depth increase from 150 to 800 nm under the surface, fully consistent with the trend shown in Figure \ref{fig:SROMapCurvatureCorrected} from Poisson-KNN analyses. Again, we see that the Sn-Sn occurrence in the first few KNN shells is slightly more favored than that in random alloys for 1NN-4NN, while getting closer to that in random alloys when K increases. These results are also consistent with the Poisson-KNN results before rearrangement.

\section{Conclusion}

In conclusion, we demonstrate a new method to retrieve 3D nanoscale SRO distribution and semi-quantitative strain distribution in GeSn alloys from APT data by Poisson-KNN statistical methods. A combination of Poisson statistics and nominal KNN analysis in APT is used to reconstruct the true KNN shells based on diamond cubic crystal structure. The 3D SRO mapping shows a relatively large fluctuation of Sn-Sn 1NN SRO parameter $\alpha^{KNN}_{Sn-Sn}$ from 0.85 to 1.15. This Sn-Sn 1NN SRO fluctuation in GeSn, together with the theoretical prediction of large band structure variation with SRO,\cite{cao2020short} contributes to the bandedge softening compared to pure Ge, as commonly observed in GeSn absorption spectra as well as  PL peak width$>>k_BT$ thermal broadening even at 10 K.\cite{olorunsola2021impact,zhou2020electrically,ye2015absorption,chen2014structural} The SRO mapping also indicates more preference towards Sn-Sn clustering near the surface, consistent with the tendency of Sn surface segregation. Furthermore, the SRO mapping shows that compressive strain tends to favor Sn-Sn 1NN, indicating that strain relaxation or tensile strain helps to prevent undesirable Sn-Sn clustering or segregation. In contrast, Sn-Sn 1NN SRO shows little dependence on the Sn composition, indicating that local strain plays a more significant role in determining the tendency of Sn-Sn clustering than the composition fluctuations from 12.5 to 15 \% Sn. Furthermore, Sn-Sn 1NN-4NN are all slightly more favored than those in random alloys, as opposed to slightly less favored Sn-Ge in 1NN-4NN shells, while their distributions get closer to that in random alloys when K increases from 1 to 4. Finally, we found that the preference of Sn-Sn 1NN is almost the same as that of 3NN, meaning that the CVD grown GeSn film is midway between random alloy and SRO structures at thermodynamic equilibrium predicted theoretically.\cite{cao2020short} This result is reasonable since CVD growth is neither an equilibrium process nor a totally random process. Furthermore, we developed an algorithm based on least square fit to rearrange atoms to nearby perfect lattice sites, which show a very good agreement in SRO analyses with the Poisson-KNN statistical results before rearrangement. 

A unique advantage of APT-based SRO analyses is that we can achieve SRO mapping and semi-quantitative strain mapping in 3D at nanoscale resolution ($\sim$10 nm) in a relatively large region with millions of atoms (e.g., 30$\times$30$\times$800 nm$^3$). It nicely bridges the gap between macroscopically averaged EXAFS analyses and atomic scale STEM analyses in studying SRO, and it can be readily extended to other material systems such as high-entropy alloys.


\section{Experimental Section}
APT samples are prepared by a FEI Helios 660 dual-beam FIB-SEM, following standard liftout technique. Fifty nm of Ni is deposited on top of the sample via sputter-coating to improve statistics in the region of interest. APT is performed on a Cameca LEAP 4000X HR instrument, with 75 pJ laser pulse energy, 100 kHz pulse frequency, 0.5\% target detection rate, and 40 K base temperature. APT data is reconstructed using Cameca IVAS software, version 3.6.10.

The Poisson-KNN method and the atomic rearrangement algorithm are implemented in Python and MATLAB. The Python package ``apt-tools'' \cite{Branson2016Git} is used to import APT data. EXAFS data are analyzed by ``Demeter'' \cite{ravel2005athena} software.

\medskip
\noindent\textbf{Supporting Information} \par 
Poisson-KNN method on epitaxial Ge on Si. Short-range order after random labelling Ge and Sn atoms. Nominal strain mapping. EXAFS measurement and analysis.

\medskip
\noindent\textbf{Acknowledgements} \par 
This research has been supported by the Air Force Office of Scientific Research under the award number FA9550-19-1-0341 managed by Dr. Gernot Pomrenke. The authors would like to thank Prof. Tianshu Li at George Washington University for theoretical input on SRO in GeSn, as well as Drs. Xiaoyi Zhang and Wenhui Hu at Argonne National Laboratory for providing the EXAFS data of GeSn used in the Supporting Information.

\medskip
\noindent\textbf{Conflict of Interest} \par 
The authors declare no conflict of interest.

\medskip
\noindent\textbf{Author Contributions} \par 
J.L. conceived the idea of Poisson-KNN method and supervised this research. S.L. further derived the mathematical equations and wrote program to implement the Poisson-KNN method and the atomic rearrangement algorithm. A.C. did random labeling and analyzed the data. X.W. did EXAFS data analysis. C.C. and W.D. contributed to APT data reconstruction. A.A. conducted sample preparation and APT data acquisition. S.Y. led materials growth and provided the GeSn samples.

\medskip

%
\bibliographystyle{MSP}
\bibliography{GeSnAPT_arXiv}



\end{document}